\documentclass[pra,twocolumn,10pt,amsmath, superscriptaddress,notitlepage,showpacs]{revtex4-1}
\usepackage{amsmath,amsfonts,amssymb,amstext,amscd,amsthm,dsfont,bbm,natbib}
\usepackage{physics}
\usepackage{color,comment}
\usepackage{soul,xcolor}
\usepackage{gensymb}
\usepackage[colorlinks=true,citecolor=violet,urlcolor=violet,linkcolor=blue]{hyperref}
\usepackage{graphicx}
\usepackage{bbold}
\usepackage{braket}
\usepackage{placeins,comment}

\begin{document}

\title{Petz recovery maps for qudit quantum channels }

\author{\ Lea Lautenbacher}
\email{lea.lautenbacher@uni-ulm.de}
\affiliation{Institut f\"ur Theoretische Physik, Albert-Einstein-Allee 11, Universit\"at Ulm, D-89069 Ulm, Germany}

\author{Vinayak Jagadish}
\email{vinayak.jagadish@helsinki.fi}
\affiliation{Instytut Fizyki Teoretycznej, Uniwersytet Jagiello{\'n}ski, {\L}ojasiewicza 11, 30-348 Krak\'ow, Poland} 
\affiliation{QTF Centre of Excellence, Department of Physics, University of Helsinki, P.O. Box 43, FI-00014 Helsinki, Finland}

\author{Francesco Petruccione}
	\affiliation{School for Data Science and Computational Thinking, Stellenbosch University, Stellenbosch 7600, South Africa}
	\affiliation{National Institute for Theoretical and Computational Sciences (NITheCS),  Stellenbosch 7600, South Africa}
 
\author{Nadja K. Bernardes}
\email{nadja.bernardes@ufpe.br}
\affiliation{Departamento de F\'{\i}sica, Universidade Federal de Pernambuco, Recife, PE  50670-901 Brasil}

\begin{abstract}
This study delves into the efficacy of the Petz recovery map within the context of two paradigmatic quantum channels: dephasing and amplitude-damping. While prior investigations have predominantly focused on qubits, our research extends this inquiry to higher-dimensional systems. We introduce a novel, state-independent framework based on the Choi-Jamio{\l}kowski isomorphism to evaluate the performance of the Petz map. By analyzing different channels and the (non-)unital nature of these processes, we emphasize the pivotal role of the reference state selection in determining the map's effectiveness. Furthermore, our analysis underscores the considerable impact of suboptimal choices on performance, prompting a broader consideration of factors such as system dimensionality.

\end{abstract}

\maketitle

\section{Introduction}
\label{Intro}

A bijective function, also known as an invertible function, refers to a map between two sets such that each element of one set corresponds to exactly one element of the other, and every element of the sets is paired. In a physical system, a similar idea appears as a reversible process related to the inversion of the direction of a process that maps the set of outputs into the set of inputs. The reversibility of a process is usually connected to thermodynamics and the variation of entropy~\cite{Norton2016}. In the quantum realm, only a unitary evolution that describes the dynamics of an isolated quantum system is reversible. However, no quantum system is truly isolated~\cite{RivasBook,haroche_exploring_2006}, but is inexorably coupled to the environment. The type of environment that affects the quantum system of interest indeed depends on the physical setting under consideration. This coupling to the environment makes the quantum system noisy. Therefore, to use quantum systems for information processing tasks, one needs to combat the effects of the interaction with the environment. This also amounts to studying whether or in what circumstances they could be undone. This naturally leads to the question of quantum recovery channels. A standard example of a recovery channel is in quantum error correction where the states in the ``code subspace" are corrected by the appropriate recovery map.

 A quantum system is associated with a Hilbert space $\mathcal{H}$, such that the set of all possible states can be defined as $\mathcal{D}(\mathcal{H}) = \{\rho \in \mathcal{L}(\mathcal{H})\}$, where $\mathcal{L}(\mathcal{H})$ represents the set of linear operators acting on $\mathcal{H}$. The most general transformations on the quantum system of interest are characterized by a family of linear maps that are completely positive and trace preserving (CPTP) known as quantum channels~\cite{Quanta77}. For certain quantum channels and input states satisfying an information theoretic criterion, a universal recovery operation was proposed by Petz~\cite{Petz86}. One should keep in mind that the recovery map should preserve the existing correlations between the system and the environment. The Petz recovery map has been widely explored in the context of quantum data processing inequality~\cite{Petz86, Petz88,Petz03,Junge18}, quantum error correction~\cite{Barnum02}, deriving fluctuation theorems~\cite{Buscemi21,Cenxin21}, to name a few. Recently, a physical protocol has been proposed to achieve this recovery map~\cite{Kwon22} and a quantum algorithm has been developed to implement it~\cite{Gilyen22}.

Although the Petz map is a powerful recovery map, it is built to perfectly work in the context where the quantum relative entropy does not increase in time. There are two crucial ingredients in constructing the Petz map: the original dynamical map (the map intended to be reverted) and a reference state (also called a prior state). For one-qubit decoherence channels the best strategy using the Petz map was studied in~\cite{Lea21}. Other general recovery strategies have been explored in~\cite{Surace22, Arthur23} based on the concept of state retrodiction.

In this work, we explore the Petz recovery map in higher dimensions. High-dimensional systems have become of increasing relevance recently \cite{Erhard18,Friis19,Erhard20,Carine20}. Two paradigmatic quantum channels are analyzed: the dephasing and amplitude-damping channels. Especially for systems based on multicore-fibers dephasing is a very typical source of error \cite{Erhard20}. Moreover, dephasing and amplitude damping channels for a qudit have been experimentally simulated in~\cite{Marques15} and the relevance and consequences of generalized amplitude damping channels have been investigated in~\cite{Khatri20,Chessa21}. Similar to what was done in Ref.~\cite{Lea21} for the qubit case, we would like to verify using the Petz map in a general context. The input states considered are the full set of states and not just the states for which this map is intended to work perfectly. By exploring the reference state, an optimal Petz map is constructed. In Ref.~\cite{Lea21}, it was shown that while the identity channel, i.e. simply returning the output state, outperforms the Petz map for unital maps in terms of average fidelity, the behavior of the Petz map for non-unital maps remains ambiguous. Since the average fidelity between a target pure state and a randomly chosen state decreases with the dimension of the systems \cite{Zyczkowski05}, we ask whether the Petz recovery map could exhibit a better performance in this general scenario with qudits. In this work, we compare the behavior of the Petz recovery map across different dimensions rather than alternative strategies, demonstrating its robustness with an increase in dimensions using a distance measure as the figure of merit. Note that no optimization over input and reference states for the Petz map was conducted.

The article is structured as follows. In Sec.~\ref{prelims}, we provide a comprehensive review of the fundamental concepts surrounding quantum channels and Petz recovery maps. Sec.~\ref{Scalability} is dedicated to exploring the behavior of the Petz map in higher dimensions and its consequent implications on the recovery strategy. Our conclusions are presented in Sec.~\ref{conclusion}. In Appendix \ref{ap:non-uniform}, we offer an illustrative example of non-uniform amplitude damping, augmenting our discussion from Sec.~\ref{Scalability}. Additionally, in Appendix \ref{ap:geom_bloch}, we delve into further geometrical features of the Petz recovery maps, providing supplementary insights to the findings in Sec.~\ref{Scalability}.

\section{Preliminaries}
\label{prelims}

\subsection{Quantum Channels}
\label{channels}

The most general transformation of a quantum system is described by completely positive and trace-preserving maps known as \emph{quantum channels}~\cite{Quanta77}.
Consider a linear map $\Lambda: \mathcal{L}(\mathcal{H}) \mapsto \mathcal{L}(\mathcal{H})$, where $\mathcal{L}(\mathcal{H})$ denotes the vector space of linear operators acting on the Hilbert space $\mathcal{H}$. We assume that ${\rm dim}\, \mathcal{H} = d$. Here, we briefly discuss a few representations of $\Lambda$ that we make use of. 
  
Fixing an orthonormal basis $\{|1\rangle, \ldots,|d\rangle\}$ in $\mathcal{H}$, one can define the operator $J(\Lambda)$, usually referred to as the Choi matrix 
\cite{Choi75,jam72},
\begin{equation}
\label{eq:choi}
  J(\Lambda) = \sum_{i,j=1}^d |i\rangle \langle j| \otimes \Lambda(|i\rangle \langle j|) ,
\end{equation}
and the corresponding $d^2 \times 
 d^2$ Hermitian  matrix
\begin{equation}
   J(\Lambda)_{ij,kl} := \langle i \otimes j|  J(\Lambda) | k \otimes l\rangle  = \langle j| \Lambda( |i\rangle \langle k|)| l\rangle .
\end{equation}
The map $\Lambda$ is completely positive if and only if $J(\Lambda) \geq 0$. 

Another useful representation is the $\mathcal {A}$ matrix obtained by reshuffling $J(\Lambda)$ as follows:
\begin{comment}
\end{comment}
\begin{equation}
  \mathcal {A}=J(\Lambda)^\mathcal{R},~~~ \mathcal {A}_{ik,jl}=J(\Lambda)_{ij,kl}.
\end{equation}
If $\Lambda$ is completely positive, it can always be represented in the operator-sum form as 
\begin{eqnarray}
  \Lambda(X) &=& \sum_i K_i X K_i^\dagger,\nonumber\\
  \sum_i K_{i}^{\dagger} K_{i} &=& \mathbb{1}\thinspace (\mathrm{Trace\, Preservation}).
\end{eqnarray}
The operators $K_i$'s are usually referred to as the Kraus operators. 
\begin{comment}
The three representations can be related as follows.
\begin{equation}
   \mathfrak{B} = \sum_i |K_i   \rangle \! \rangle  \langle \! \langle K_i|
\end{equation}
and
\begin{equation}
\mathcal{A} =  \sum_i K_i \otimes K^{*}_i.
\end{equation}
\end{comment}

There exists another representation referred to as the Bloch representation. To this end, let $\Gamma_i$ $(i=0,1,\ldots,d^2-1)$ denote a Hermitian orthonormal basis in $\mathcal{L}(\mathcal{H})$, ${\rm Tr}(\Gamma_i \Gamma_j) = \delta_{ij}$, and let $\Gamma_0 = \mathbb{1}/\sqrt{d}$. The map acting on a general density matrix $\rho$ of a qudit can be represented as  
\begin{equation}
    \label{eq:rho_qubit}
     \Lambda(\rho) = \Lambda\Big(\frac{\mathbb{1} + \vec{r}.\vec{\Gamma}}{d}\Big)= \frac{\mathbb{1} + (M\vec{r}+\vec{\tau}).\vec{\Gamma}}{d} ,
\end{equation}
where $\vec{\tau} \in \mathbb{R}^{d^2-1}$, $M$ is
a real square matrix of order 
$d^2-1$ and $\vec{r}$ is referred to as the Bloch-vector corresponding to the qudit ~\cite{Kimura, Krammer}. This represents the transformation of the Bloch vector, $\vec{r} \to (M\vec{r}+\vec{\tau}) $, and is commonly known as the \emph{affine } representation.

A quantum channel is said to be \emph{unital} if the identity is preserved. i.e., $\Lambda(\mathbb{1}) = \mathbb{1}$. 

The \emph{dual map}, $\Lambda^{\dagger}$, in the operator sum representation is defined as \begin{equation}
  \Lambda^\dagger(X) = \sum_i  K_i^\dagger X K_i.
\end{equation}

Let us now briefly discuss two channels that we analyze in this work.

 \subsubsection{Dephasing Channels}

The dephasing channel is one of the most common decoherence channels.  It is an unital map that destroys the relative phases between the computational basis states, $\{|k\rangle\}$.
The uniform dephasing map is defined as
\begin{equation}
\label{eq:dephasing}
    \Lambda_{D}(\rho)= (1-p)\rho+ p\sum_{k=0}^{d-1}P_k\rho P_k,
\end{equation}
with the projector $P_k=|k\rangle\langle k|$. The parameter $p\in[0,1]$ characterizes the noise strength ($p=0$, no noise; $p=1$ maximal noise). For $p=1$, the coherences vanish completely.

\subsubsection{Damping Channels}

Damping channels encompass phenomena such as amplitude damping, phase damping, and thermal relaxation, each characterized by different mechanisms of energy dissipation. While specific channel models may vary, damping channels generally involve the transition of a quantum system from an excited state to a lower energy state, often due to interactions with the surrounding environment. Here, we consider the amplitude damping map $\Lambda_{AD}$ defined as
\begin{align}
\label{eq:KrausAD}
\Lambda_{AD}(\rho)&= \sum_{i=0}^{d-1} K_{i}\rho K_{i}^\dagger, \mathrm{with}\nonumber \\
    K_{0} &= |0\rangle\langle 0| + \sqrt{1-p}\sum_{i=1}^{d-1} |i\rangle\langle i|\nonumber \\
    K_{i=1..d-1} &= \sqrt{p} |0\rangle\langle i|.
\end{align}
In this specific representation, we model the damping to the ground state $\ket{0}$ for a qudit system of dimension $d$, where $\{\ket{i}\}_{i = 0,1,2,\cdots,d-1}$ denotes an orthonormal basis of the Hilbert space of the qudit. For simplicity, here we include the transitions from other levels to the state $\ket{0}$ alone and ignore the transitions between other levels. One could also have damping maps with varying $p$. An example for the qutrit case is discussed in Appendix~\ref{ap:non-uniform}.

\subsection{Non-unitality measure}
\label{NUnital}
Another property we take into account is the non-unitality of the map. Consider a quantum channel $\Lambda$, such that $\Lambda(\mathbb{1}_d/d) = \rho'$, with $\mathbb{1}_d/d$ the maximally mixed state for a qudit with dimension $d$. We define a non-unitality measure $\mathcal{N}$ as the deviation of the output state $\rho'$ from the maximally mixed state. Employing state distinguishability, we write
\begin{equation}
    \label{eq:NUmeasure}
    \mathcal{N} = \frac{1}{2} ||\rho' - \frac{\mathbb{1}_d}{d}||_1, 
\end{equation}
where $||A||_1 = \text{Tr}\{\sqrt{AA^{\dagger}}\}$ and $\mathcal{N}=0$ iff $\Lambda$ is unital. This way of estimating the non-unitality of the map is convenient, especially for higher dimensions.

\subsection{Petz Recovery Maps}
\label{Recovery}

An important definition we use in this work is the recoverability property of a map. We say a map is \textit{recoverable} if there exists another CPTP map $\Tilde{\Lambda}$ such that 
\begin{equation}
    \label{eq:recovery}
    \Tilde{\Lambda}(\Lambda(\rho)) = \rho,\, \forall\rho\in\mathcal{S}(\mathcal{H}).  
\end{equation}
The map $\Tilde{\Lambda}$ is known as the recovery map of $\Lambda$ for the subset $\mathcal{S}(\mathcal{H})$, such that $\mathcal{S}(\mathcal{H}) \subseteq \mathcal{D}(\mathcal{H})$. The most well-known class of recovery maps are the Petz recovery maps~\cite{Petz86, Petz88, Petz03}, defined as 
\begin{equation}
    \label{eq:petz}
    \mathcal{P}^{\sigma}(\rho) = \sigma^{\frac{1}{2}} \Lambda^\dagger\left(\Lambda(\sigma)^
{-\frac{1}{2}}\;\rho\;\Lambda(\sigma)^{-\frac{1}{2}}\right)\sigma^{\frac{1}{2}},
\end{equation}
with $\Lambda^\dagger$ the dual of $\Lambda$ and $\sigma$ the reference state, such that $\text{supp}(\rho) \subseteq \text{supp}(\sigma)$. In the process of recovery, all states in $\mathcal{S}(\mathcal{H})$ should be treated equally. This imposes the introduction of a fiducial reference state $\sigma$, without focusing on a particular state that one is interested in, say $\rho$.  The reference state, therefore, becomes handy to determine whether the map $\Lambda$ affects the relative entropy between the states $\sigma$ and $\rho$. The subset where a certain map is recoverable is defined by the invariability of the relative entropy after the action of the map, i.e, 
\begin{equation}
\label{eq:entropy}
    S(\rho||\sigma) =
S(\Lambda(\rho)||\Lambda(\sigma)), \,\forall \rho,\sigma\in \mathcal{S}(\mathcal{H}), 
\end{equation}
with $S(\rho||\sigma) = \text{Tr}(\rho\log\rho - \rho\log\sigma)$ being the relative entropy between two states. If Eq. (\ref{eq:entropy}) is satisfied, then there exists a state $\sigma$ such that $\mathcal{P}^{\sigma}( \Lambda(\rho))=\rho$.

Let us now represent the Petz map in the Choi matrix form. Given a $n\times n$ matrix $X$, $|X \rangle\rangle$ denotes the vectorized version of $X$ where the matrix is row-wise folded into a column vector. Let us express the Petz map acting on the vectorized form of the density matrix, i.e. $|\mathcal{P}^{\sigma}(\rho)\rangle\rangle = \mathcal{A}_{\text{Petz}} |\rho\rangle\rangle $. In order to do it,  from Eq.~(\ref{eq:petz}), we expand $ \Lambda^\dagger\left(\Lambda(\sigma)^{-\frac{1}{2}}\;\rho\;\Lambda(\sigma)^{-\frac{1}{2}}\right)$ as follows
\begin{align*}
    \Lambda^\dagger\left(\Lambda(\sigma)^{-\frac{1}{2}}\;\rho\;\Lambda(\sigma)^{-\frac{1}{2}}\right)&= \Lambda^\dagger\left(\sigma'\;\rho\;\sigma'\right)\\
    &= \sum_{\alpha}K_{\alpha}^{\dagger}\left(\sigma'\;\rho\;\sigma'\right)K_{\alpha} \\
    &= \sum_{\alpha}M_{\alpha} \rho N_{\alpha},
\end{align*}
with $\sigma' = \Lambda(\sigma)^{-\frac{1}{2}}$, $M_\alpha = K_{\alpha}^{\dagger}\sigma'$, and $N_{\alpha} = \sigma'K_{\alpha}$. Note that the dual map $\Lambda^\dagger$ being completely positive, admits an operator-sum representation, with the Kraus operators $\{K_{\alpha}^{\dagger}\}$. Using the identity $|X.Y.Z\rangle\rangle = X \otimes Z^{T}|Y\rangle\rangle$~\cite{Quanta77,KarolBook} ($X,Y,Z$ being arbitrary square matrices) twice, we obtain
\begin{equation}
    \mathcal{A}_{\text{Petz}} |\rho\rangle\rangle = \Big(\sigma^{\frac{1}{2}} \otimes (\sigma^{\frac{1}{2}})^T . \sum_\alpha M_{\alpha} \otimes N_{\alpha}^{T} \Big)|\rho\rangle\rangle. 
\end{equation}
The corresponding Choi matrix can be obtained by reshuffling the matrix $\mathcal{A}_{\text{Petz}}$.

\section{Scalability of the Petz maps}
\label{Scalability}

As the dimensionality of the quantum system increases, the complexity of quantum operations and state reconstruction also increases. In higher dimensions, the space of quantum states becomes exponentially larger, leading to greater challenges in quantum information processing tasks such as state tomography and error correction. Contrary to what has been done in \cite{Lea21}, the average fidelity now is not numerically efficient for higher dimensional systems. For this reason, we propose a state-independent figure of merit as a quantifier of the performance of the Petz recovery map. Ideally, if the Petz map is capable of recovering the initial state $\rho$, we would obtain $\mathcal{P}^{\sigma}(\Lambda (\rho)) =  \rho$, which implies
\begin{equation}
    \label{eq:petz_inv}
    \mathcal{P}^{\sigma}\circ\Lambda =  \mathcal{I}, 
\end{equation}
where $\mathcal{I}$ is the identity channel. 
In order to check the validity of Eq.~(\ref{eq:petz_inv}), we propose a measure of \textit{distance} between maps, given by 
\begin{equation}
\label{eq:dist_inverse}
D(\mathcal{P}^{\sigma}\circ\Lambda, \mathcal{I}) = ||J(\mathcal{P}^{\sigma}\circ\Lambda) - J(\mathcal{I}) ||_1,
\end{equation}
where $J(\Lambda)$ is the Choi matrix corresponding to $\Lambda$. A map $\Lambda$ has an inverse $\Lambda^{-1}$ if $\Lambda^{-1}(\Lambda(\rho)) = \rho,\, \forall\rho\in \mathcal{D}(\mathcal{H})$. The measure above reflects how far the Petz map is from the inverse of the map $\Lambda^{-1}$. If $\mathcal{P}^{\sigma} = \Lambda^{-1}$, we would expect to obtain $D(\mathcal{P}^{\sigma}\circ\Lambda,\mathcal{I}) = 0$. The inverse map is usually not completely positive unless the original map is a unitary~\cite{jagadish_measure2_2019}. 

The measure above is well-behaved and bounded by~\cite{Wallman}
\begin{equation}
\label{eq:bounds}
    \frac{1}{d} ||\Lambda_1 - \Lambda_2||_\diamond \leq ||J(\Lambda_1) - J(\Lambda_2) ||_1 \leq ||\Lambda_1 - \Lambda_2||_\diamond,
\end{equation}
with $||\Lambda||_\diamond \equiv \text{max}_{\rho} ||(\mathbb{1}\otimes\Lambda)\rho||_{1}$ being the diamond norm. The channels are perfectly distinguishable if $||\Lambda_1 - \Lambda_2||_\diamond = 2$. Observe that the measure proposed in Eq.~(\ref{eq:dist_inverse}) can be more easily calculated, since it does not require an optimization as the diamond norm. In order to obey the bound in Eq.~(\ref{eq:bounds}), we  normalize the Choi matrix in Eq.~(\ref{eq:choi}) by the factor $1/d$. Now we have a state-independent framework, which is quite useful and has a lower computational cost when increasing the dimension. To obtain the Petz recovery map, we parametrize the reference state as 
\begin{equation}
    \label{eq:sigma_dim}
    \sigma(\epsilon) = (1 - \epsilon) |0\rangle\langle 0| + \frac{\epsilon}{d-1}  \sum_{n=1}^{d}|n\rangle\langle n|, 
\end{equation}
with $\epsilon \in [0,1]$. This allows us moving to the $d$-dimensional scenario. 
\begin{figure}[h]
    \centering
\includegraphics[width=0.48\textwidth]{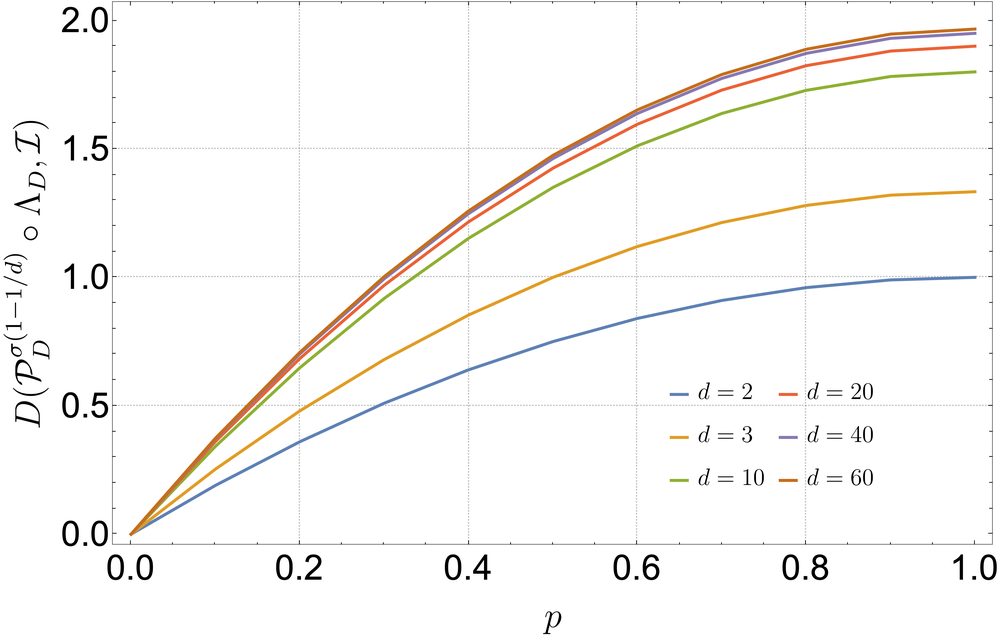}
    \caption{Distance between the Petz map applied to the dephasing channel and the identity channel, $D(\mathcal{P}_D^{\sigma(1 - 1/d)}\circ\Lambda_D, \mathcal{I})$, versus the noise strength $p$ for different dimensions.}
    \label{fig:DistXpDim}
\end{figure}

\begin{figure}[h]
    \centering
\includegraphics[width=0.48\textwidth]{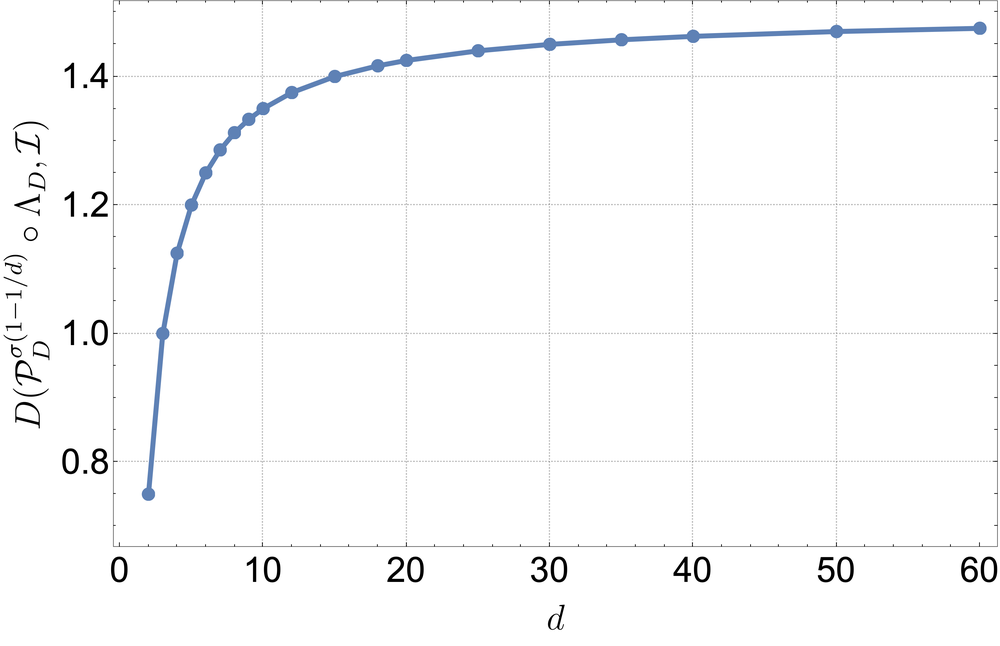}
    \caption{Distance between the Petz map applied to the dephasing channel and the identity channel, $D(\mathcal{P}_D^{\sigma(1 - 1/d)}\circ\Lambda_D, \mathcal{I})$, versus the dimension $d$ for fixed $p = 0.5$.}
    \label{fig:distXdim}
\end{figure}

For the dephasing channel, we fix the reference state as the maximally mixed one, $\sigma(1 - 1/d)$. A plot of the distance $D(\mathcal{P}_D^{\sigma(1 - 1/d)}\circ\Lambda_D, \mathcal{I})$ versus $p$ for different dimensions $d$ is shown in Fig.~\ref{fig:DistXpDim} . We observe that the displacement between the curves decreases when increasing the dimension and gets much smaller as we move towards $\text{max} \, D(\mathcal{P}_D^{\sigma(1 - 1/d)}\circ\Lambda_D, \mathcal{I}) = 2$. In order to obtain a better relation between recovery and dimension, in Fig.~\ref{fig:distXdim} we plot $D(\mathcal{P}_D^{\sigma(1 - 1/d)}\circ\Lambda_D, \mathcal{I})$ versus the dimension fixing $p = 0.5$. As can be observed, the distance between the two maps stabilizes when increasing the dimension. As an example, taking $d = 20$ we obtain $D(\mathcal{P}_D^{\sigma(1 - 1/d)}\circ\Lambda_D, \mathcal{I}) = 1.42$, for $d = 40$, $D(\mathcal{P}_D^{\sigma(1 - 1/d)}\circ\Lambda_D, \mathcal{I}) = 1.46$ and for $d = 60$, $D(\mathcal{P}_D^{\sigma(1 - 1/d)}\circ\Lambda_D, \mathcal{I}) = 1.47$. We observed that for the chosen reference state, the Petz map performs better in the qubit case compared to $d>2$. However, there is a certain threshold, beyond which a similar performance is observed for $d>20$. Despite the poor performance in higher dimensions compared to qubits, the relatively consistent behavior observed for different high-dimensional cases suggests a certain level of robustness of Petz recovery maps to dimensionality beyond a certain point. To support this analysis we provide further details in Appendix ~\ref{ap:geom_bloch}. It must also be noted that the depolarizing channel leads to similar conclusions as the dephasing channel, but we omit the details for brevity.

\begin{figure}[h]
    \centering
\includegraphics[width=.48\textwidth]{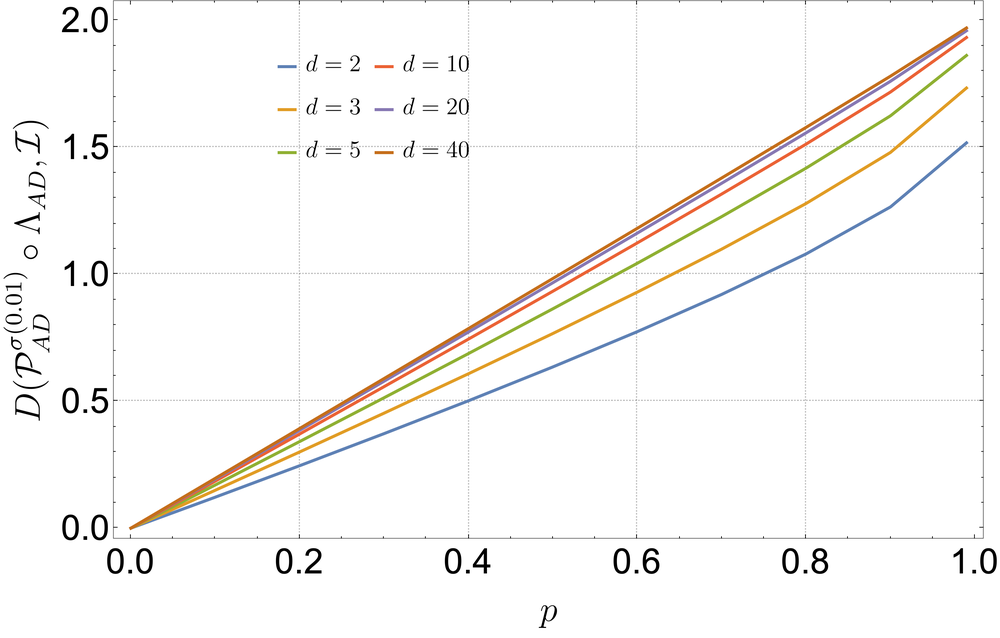}\hspace{0.5cm}
    \caption{Distance between the Petz map applied to the amplitude-damping channel and the identity channel, $D(\mathcal{P}_{AD}^{\sigma(0.01)}\circ\Lambda_{AD}, \mathcal{I})$, versus the noise strength $p$ for different dimensions.} 
    \label{fig:DistanceAD}
\end{figure}

\begin{figure}[h]
    \centering
\includegraphics[width=.48\textwidth]{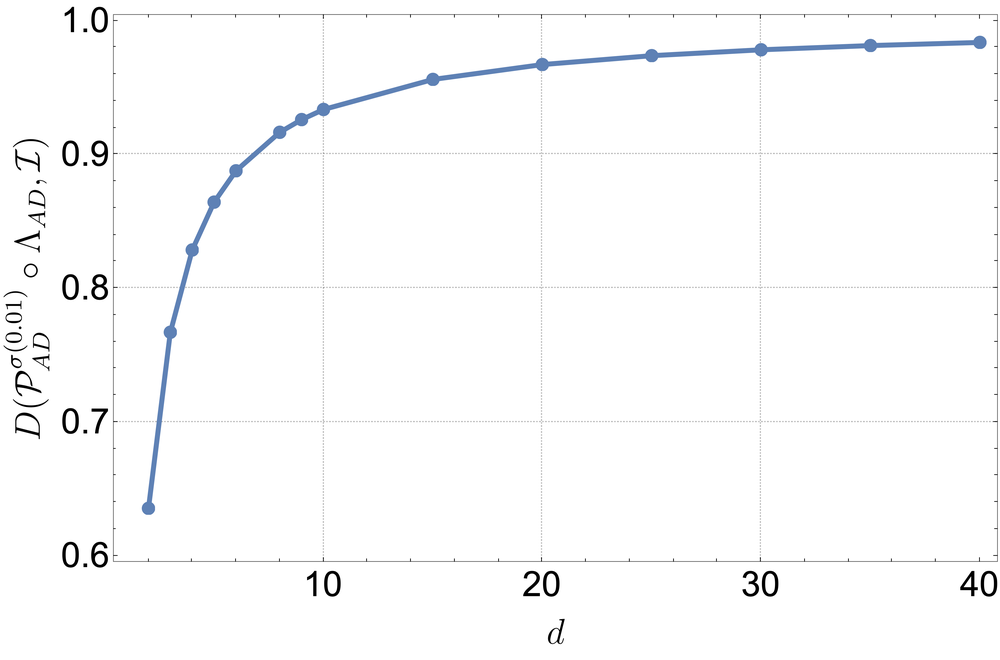}
    \caption{Distance between the Petz map applied to the amplitude-damping channel and the identity channel, $D(\mathcal{P}_{AD}^{\sigma(0.01)}\circ\Lambda_{AD}, \mathcal{I})$, versus the dimension $d$ for fixed $p = 0.5$.} 
    \label{fig:distXdimAD}
\end{figure}

\begin{figure}[t]
    \centering  
\includegraphics[width=.48\textwidth]{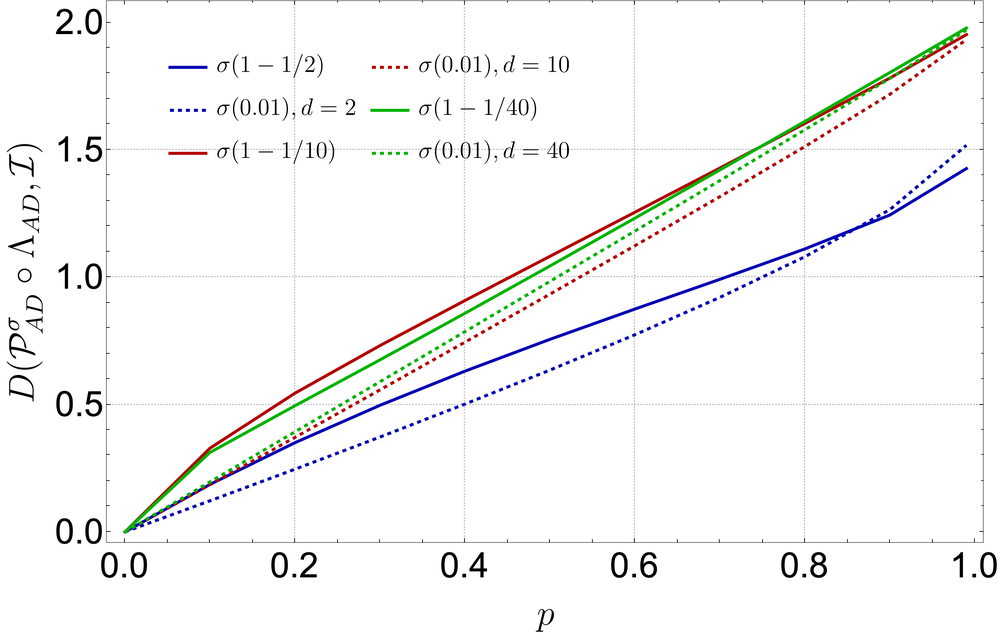}\\
\includegraphics[width=.48\textwidth]{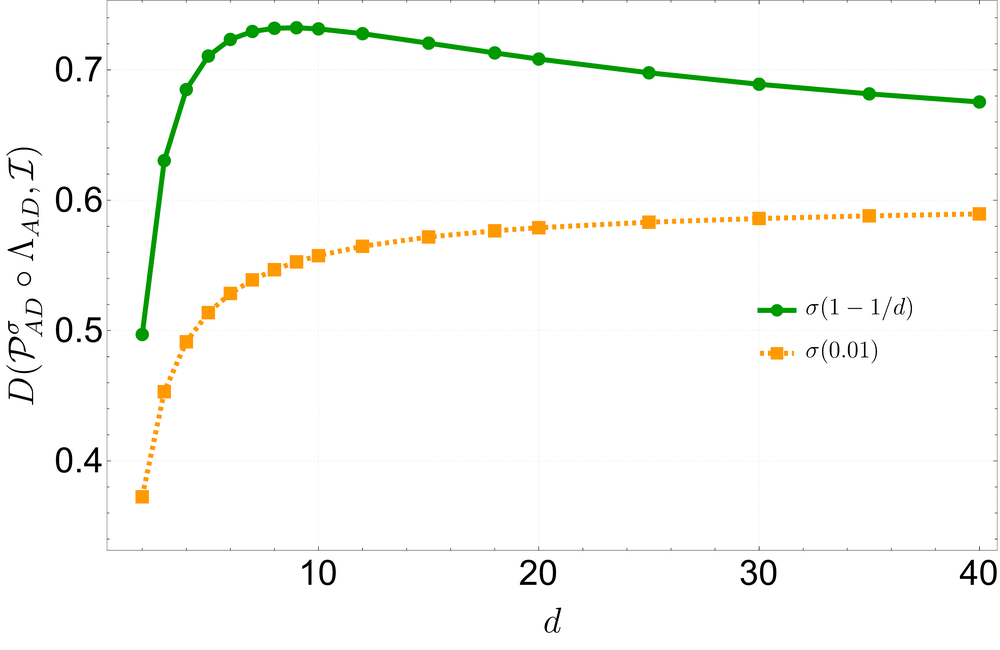}
    \caption{Distance between the Petz map applied to the amplitude-damping channel and the identity channel, $D(\mathcal{P}_{AD}^{\sigma}\circ\Lambda_{AD}, \mathcal{I})$. On top, the difference between the choice of the reference state was plotted for dimensions $d = 2, 10, \text{and } 40$ in terms of noise parameter $p$. For each dimension, the reference state analyzed is the corresponding maximally mixed state, $\sigma(1-1/d)$, and a reference state close to the fixed point of the map, $\sigma(0.01)$. On the bottom, $D(\mathcal{P}_{AD}^{\sigma}\circ\Lambda_{AD}, \mathcal{I})$ for $p=0.3$ versus the dimension $d$.}
    \label{fig:DistDimAD03}
\end{figure}

For the amplitude damping channel, a similar behavior is observed. Initially, we decided to fix the reference state close to the fixed point, $\sigma(0.01)$ \cite{footnote_1}. In Fig.~\ref{fig:DistanceAD} we observe how the distance $D(\mathcal{P}_{AD}^{\sigma(0.01)}\circ\Lambda_{AD}, \mathcal{I})$ increases with the dimension. Again, the curves are getting closer for higher dimensions. This can be easily seen in Fig.~\ref{fig:distXdimAD} for $p = 0.5$. By fitting the curve displayed in Fig.~\ref{fig:DistanceAD} and finding the continuous function that interpolates the points, we evaluate the derivative with respect to the dimension. We observe its decrease as the dimension increases. For $d \geq 20$, the derivative tends towards zero and is of the order of $10^{-2}$. Consequently, the distance $D(\mathcal{P}_{AD}^{\sigma(0.01)}\circ\Lambda_{AD}, \mathcal{I})$ transitions into a plateau. There is no difference in the recovery performance of the Petz map for $d\geq 20$.

For the sake of completeness, we include the analysis of the reference state as being the maximally mixed one $\sigma(1 - 1/d)$. In Fig.~\ref{fig:DistDimAD03} (top) we observe how the green and red solid lines present crossing points depending on the noise strength $p$, also, how the choice of reference state can significantly impact the quality of state recovery using the Petz map. For intermediate noise strengths, we can identify an advantage in the Petz map performance for $d = 40$ over $d = 10$. In Fig.~\ref{fig:DistDimAD03} (bottom), we plot the distance versus the dimension for different choices of the reference state. We see that for an intermediate noise strength, fixing the value $p = 0.3$, the advantage of choosing $\sigma$ close to the fixed point of the map is clear. However, when choosing $\sigma$ as the maximally mixed state, dimension comes into play and impacts the recovery performance. We obtain for $d=10$ and $d=40$, $D(\mathcal{P}_{AD}^{\sigma(0.5)}\circ\Lambda_{AD}, \mathcal{I}) = 0.73$ and $D(\mathcal{P}_{AD}^{\sigma(0.5)}\circ\Lambda_{AD}, \mathcal{I}) = 0.68$, respectively. Since, such behavior is not observed for the dephasing and depolarizing channels, i.e. unital maps, we expect that delving into the non-unital character of the amplitude damping channel may shed light on this issue.

\begin{figure}[h]
    \centering
\includegraphics[width=0.99\linewidth]{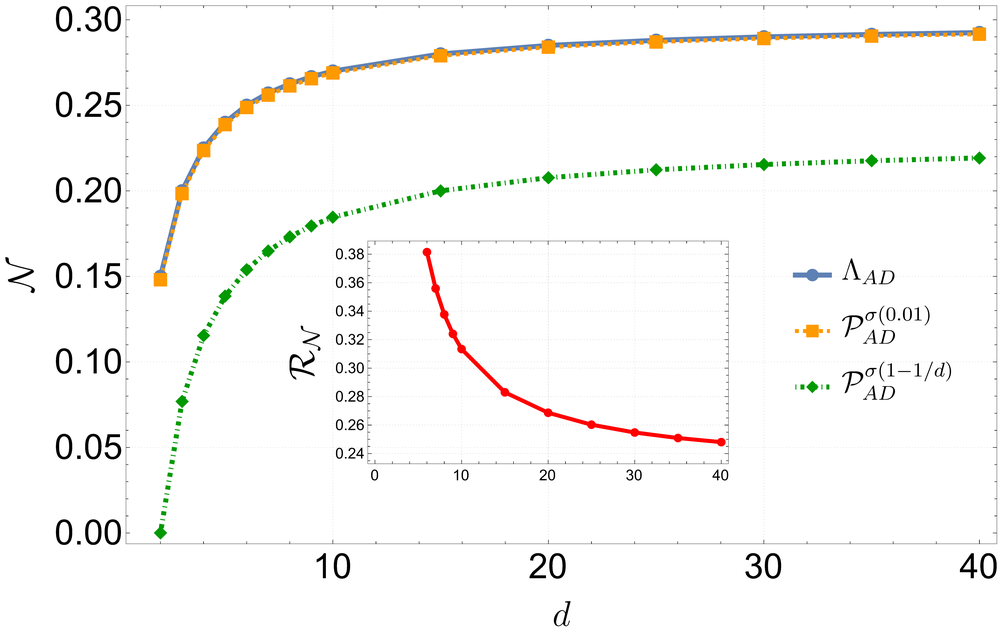}
    \caption{Non-unitality measure $\mathcal{N}$ in terms of the dimension, fixing $p = 0.3$, for the amplitude-damping channel (thick) and its corresponding Petz map for $\sigma(0.01)$ (dashed) and $\sigma(1-1/d)$ (dot-dashed). In the inset, we displayed the ratio between the non-unital character of $\mathcal{P}_{AD}^{\sigma(1 - 1/d)}$ and $\Lambda_{AD}$ in terms of the dimension.}
    \label{fig:NUnitalityAD}
\end{figure}

In Fig.~\ref{fig:NUnitalityAD} we compute the non-unitality measure, $\mathcal{N}$, introduced in Sec.~\ref{prelims} for a fixed value of $p$. For $p = 0.3$, the non-unitality measure $\mathcal{N}$ for the Petz map increases with the dimension but the outcome is different depending on the choice of the reference state. We observe that the curves for $\mathcal{P}_{AD}^{\sigma(0.01)}$ and $\Lambda_{AD}$ are superimposed, meaning that the Petz map for the choice of $\sigma(0.01)$ shifts the maximally mixed state in the same way as the amplitude-damping channel. On the other hand, for $\mathcal{P}_{AD}^{\sigma(1 - 1/d)}$, $\mathcal{N}$ is much smaller and it does not reproduce the same non-unital behavior as $\Lambda_{AD}$. Changing the noise strength $p$, the same behavior of $\mathcal{N}$ is observed but with different values. For a qubit, from Fig. \ref{fig:NUnitalityAD}, we obtain $\mathcal{N} = 0$ and $\mathcal{N} = 0.15$ for the choices of $\sigma(1 - 1/d)$ and $\sigma(0.01)$, respectively. Bringing together Figs. \ref{fig:DistDimAD03} and \ref{fig:NUnitalityAD}, a connection can be traced. When choosing $\sigma(1 - 1/d)$, which is not the optimal reference state when considering the full state space ~\cite{Lea21}, the Petz map does not retrieve the non-unital behavior of the amplitude damping, contrary to what is observed when choosing the optimal $\sigma(0.01)$. This is a limitation encoded in the choice of reference state that reflects on the bad performance of $\mathcal{P}_{AD}^{\sigma(1 - 1/d)}$ compared to $\mathcal{P}_{AD}^{\sigma(0.01)}$. In Appendix ~\ref{ap:geom_bloch} we provide additional geometric evidence supporting the choice of $\sigma(0.01)$ as the optimal reference state within states of the form of Eq.~(\ref{eq:sigma_dim}). It can be also verified that when increasing the dimension, the difference between the non-unital character of $\Lambda_{AD}$ and $\mathcal{P}_{AD}^{\sigma(1-1/d)}$ decreases. In the inset of Fig.~\ref{fig:NUnitalityAD}, we depict the ratio between the non-unitality of both maps is, given by $\mathcal{R}_{\mathcal{N}} = 1 - \mathcal{N}(\mathcal{P}_{AD}^{\sigma(1 - 1/d)})/\mathcal{N}(\Lambda_{AD})$. As one can see, the non-unital character of the map is better captured for the Petz for higher dimensions, which explains the advantage observed in Fig.~\ref{fig:DistDimAD03}. Our findings highlight the critical influence of the choice of the reference state on the performance of the Petz map. Notably, our analysis reveals that the choice of $\sigma$, as depicted in Fig.~\ref{fig:DistDimAD03}, can significantly impact the performance, prompting the consideration of additional factors such as dimensionality.

\section{Conclusions}
\label{conclusion}

In the present work, we extend what has been investigated so far in the literature for the Petz recovery map to high-dimensional quantum systems. Our focus lies on analyzing quantum channels with direct experimental implications, such as the dephasing and amplitude-damping channels. The utility of qudits stems from their ability to encode and process larger amounts of information compared to qubits, an advantage in processing tasks. However, dealing with high-dimensional systems imposes an additional challenge due to the increase of the susceptibility to environmental disturbances and errors in control operations. Managing and mitigating these sources of noise and decoherence becomes increasingly challenging as the size of the quantum state space grows. Designing efficient error correction schemes capable of protecting high-dimensional quantum states against noise and errors is an ongoing focus of research that motivates our endeavor to provide a comprehensive characterization of the Petz maps in the high-dimensional regime.

We have devised a state-independent framework that obviates the need for optimization algorithms, leveraging a distance measure based on the Choi-Jamio{\l}kowski isomorphism. Our findings emphasize the importance of carefully selecting the reference state for the Petz map. In this work, we provide numerous numerical evidences supporting the fixed point of the map as the optimal choice of the reference state. Moreover, we have discerned that for non-unital maps, the choice of reference state significantly impacts the non-unital character of the corresponding Petz map  — an essential consideration for accurately reversing the action of the channel. Notably, when exploring a broader state space, we have uncovered an intriguing phenomenon: depending on the chosen reference state, dimensionality emerges as a factor influencing the performance of the Petz map.

Our work provides a novel approach to understanding and characterizing the Petz recovery maps, that has the potential to improve quantum information processing protocols. We anticipate that the insights gained about the characteristics and performance of Petz recovery maps can provide valuable guidance in designing quantum error correction protocols that are more effective and applicable to a wide range of quantum systems.

\acknowledgements

L.L. acknowledges financial support from QuantERA project ExtraQT. V.J. acknowledges financial support from the Foundation for Polish Science
through TEAM-NET project (contract no. POIR.04.04.00-00-17C1/18-00). N.K.B. acknowledges financial support from CNPq Brazil (Universal Grant No. 406499/2021-7) and FAPESP (Grant 2021/06035-0). N.K.B. is part of the Brazilian National Institute for Quantum Information (INCT Grant 465469/2014-0). F.P. acknowledges support of the NICIS (National Integrated Cyber Infrastructure System) e-research grant QICSA and of the South African Quantum Technology Inititive (SA QuTI).
\vspace{-3 mm}

\begin{appendix}

\section{Non-uniform amplitude damping channel}
\label{ap:non-uniform}

We also investigate if changes in the recoverability can be observed in the case of non-uniform amplitude damping, i.e. when different damping rates are considered.
\begin{figure}[h]
    \centering
\includegraphics[width=0.48\textwidth]{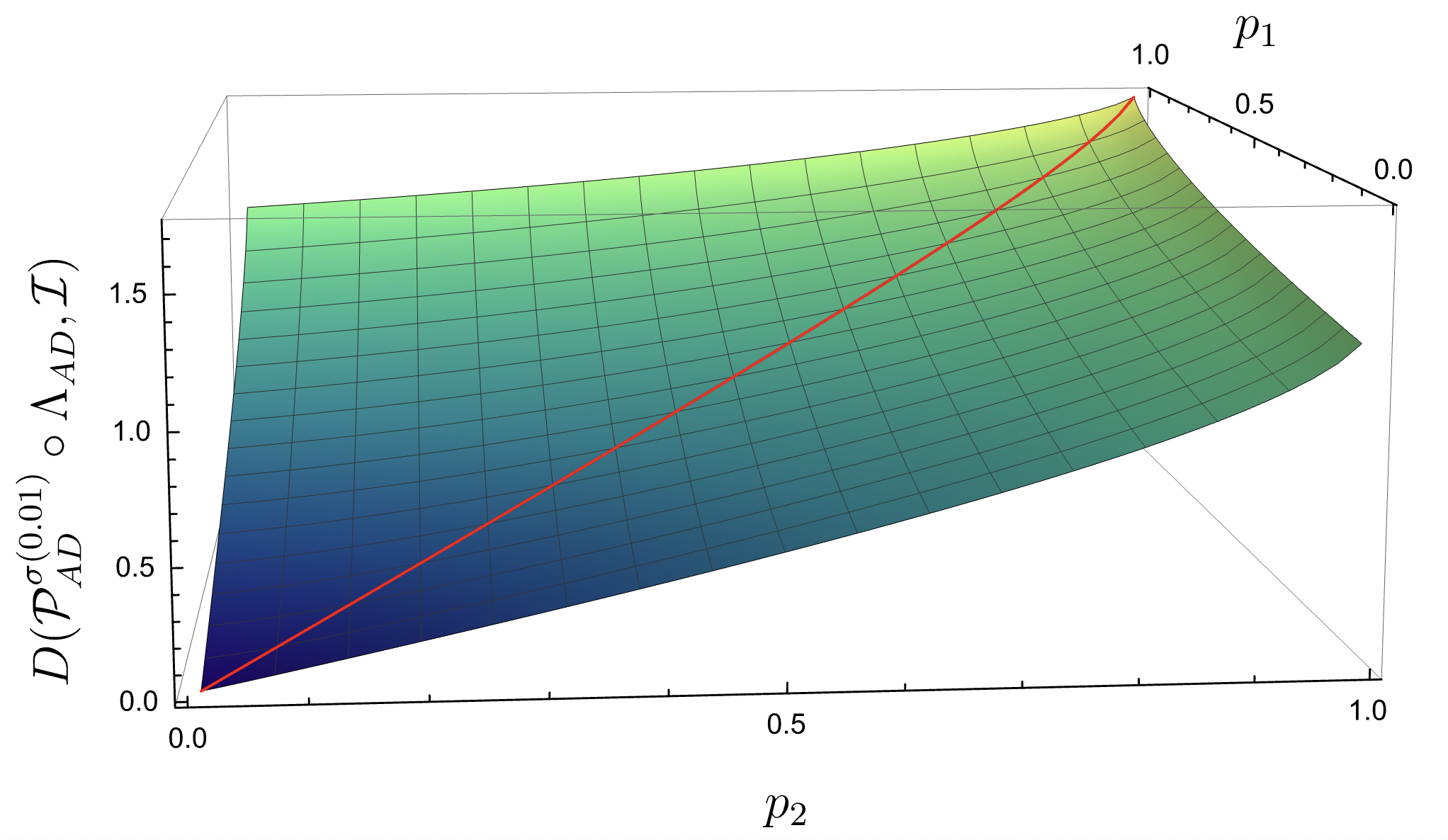}
    \caption{Distance between the Petz map applied to the amplitude-damping and the identity channel in terms of the decay probabilities $p_1$ and $p_2$. The red curve corresponds to the uniform case, $p_1 = p_2$. }
    \label{fig:UandNUAD}
\end{figure}
Let us take the qutrit case as an example. The Kraus operators are known from Eq.~(\ref{eq:KrausAD}), but now with a slight modification; 
\begin{eqnarray}
\label{eq2:KrausAD}
    K_{0} &=& |0\rangle\langle 0| + \sqrt{1-p_1}|1\rangle\langle 1| +  \sqrt{1-p_2}|2\rangle\langle 2|\nonumber\\
    K_{1} &=& \sqrt{p_1} |0\rangle\langle 1| \nonumber\\
    K_{2} &=& \sqrt{p_2} |0\rangle\langle 2|,
\end{eqnarray}
with $p_1,p_2 \in [0,1]$ being the decay probabilities. In the non-uniform case, these probabilities assume different values, which allows us to deal with a more realistic scenario. In Fig.~\ref{fig:UandNUAD} we plot the behavior of the distance defined in Eq.~(\ref{eq:dist_inverse}), in terms of $p_1$ and $p_2$. 

We can see that the distance for the uniform amplitude damping (diagonal red curve) turns out to be smaller than for the non-uniform scenario. When the decay probabilities are equal, a better recovery is observed for the Petz map. Assuming that one of the probabilities is zero, the system reduces to a two-level system, and we recover the result displayed in Fig.~\ref{fig:DistanceAD} of the main text for $d = 2$.

\section{Geometrical aspects of the Petz recovery map}
\label{ap:geom_bloch}

Here, we delve into certain aspects of the Petz recovery map that remain less explored. Specifically, we focus on discerning the characteristics of states that are recoverable within a defined subset as outlined by Eq.~(\ref{eq:entropy}). Our investigation aims to shed light on how these subsets evolve when subjected to the influence of quantum channels. To achieve this, we start from Eq.~(\ref{eq:rho_qubit}) to analyze the geometrical properties of the aforementioned quantum operations. By examining the dynamics of the generalized Bloch vector under the influence of a quantum channel and its corresponding Petz recovery map, we can delineate the \textit{accessible state space}, i.e. resultant set of states under a specific transformation. From Eq. (\ref{eq:rho_qubit}), one can note that the action of a CPTP map $\Lambda$ either rotates or shrinks the Bloch vector for unital maps along with a possible translation in the case of non-unital maps. Such transformations induce a change in the volume of the set of accessible states, defined as 
\begin{equation}
\label{eq:volume}
    V = |M|, 
\end{equation}
we indicate with $|M|$ the determinant of the matrix $M$ which carries information about the affine transformation induced on the Bloch vector $\vec{r}$ induced by $\Lambda$. The determinant of this matrix is proportional to the volume of the ellipsoid defined by $\vec{r}$. A similar analysis can also be found in~ \cite{Jevtic}. 

\begin{figure}[h]
    \centering
\includegraphics[width=0.48\textwidth]{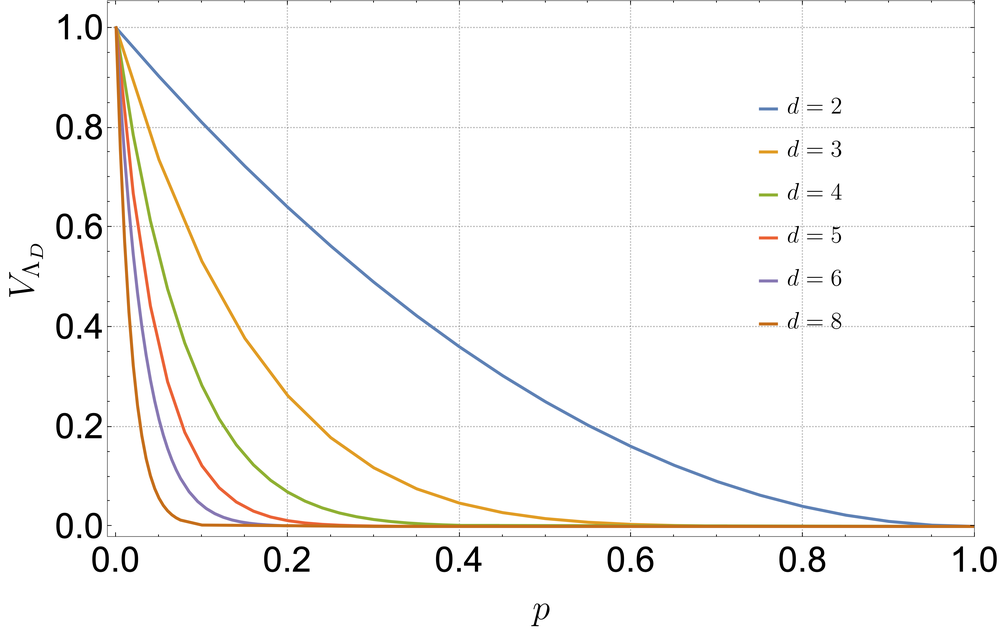}\\
\includegraphics[width=0.48\textwidth]{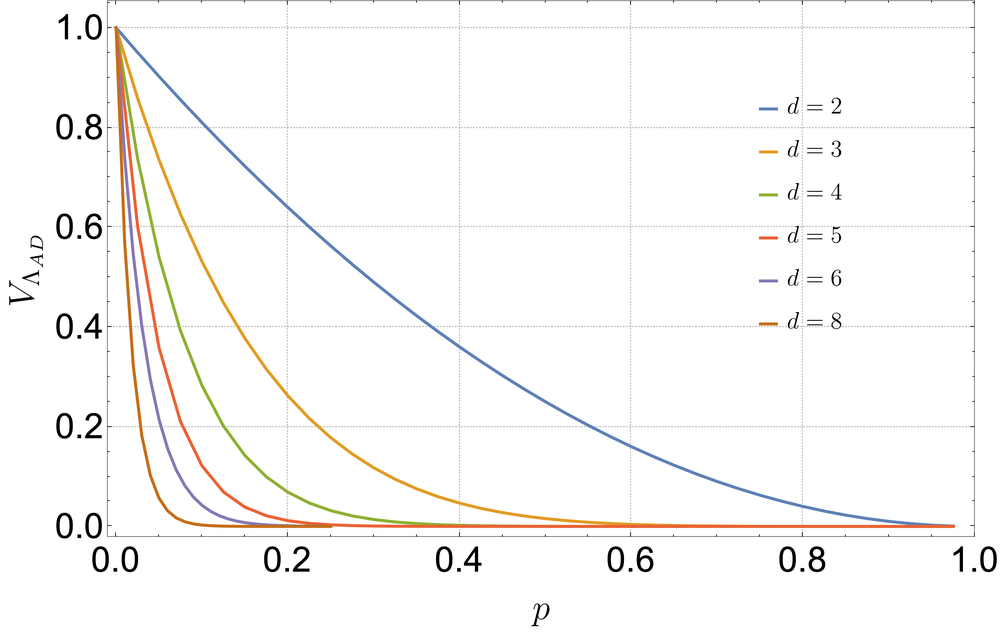}
    \caption{Volume of the set of accessible states under the dephasing map $\Lambda_D$ (top) and the amplitude damping map $\Lambda_{AD}$ (bottom) for different dimensions in terms of the noise strength $p$.}
    \label{fig:VolD}
\end{figure}

\begin{figure}[h]
    \centering
\includegraphics[width=0.48\textwidth]{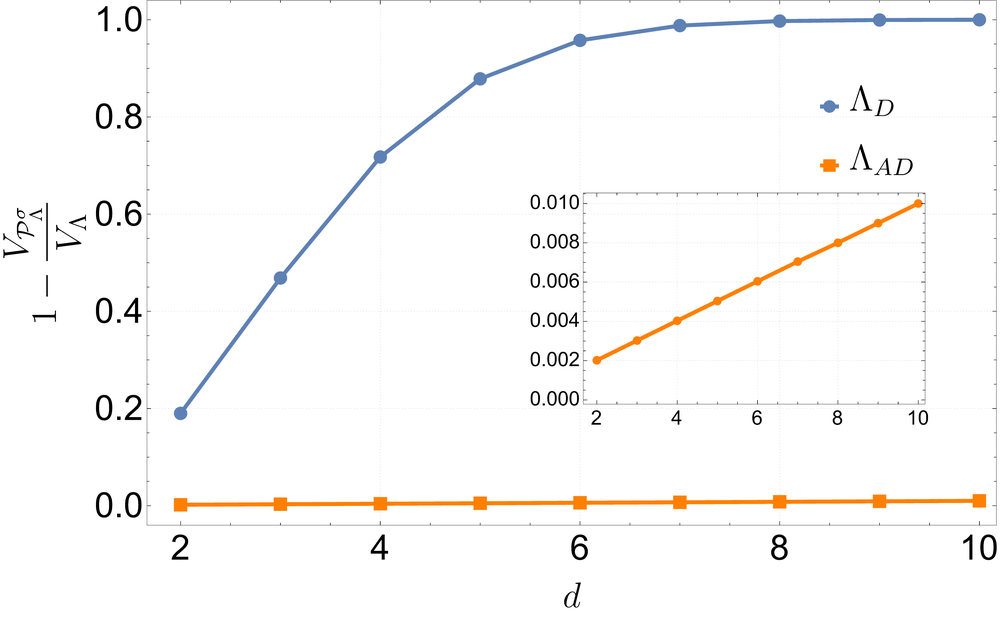}\\
    \caption{Relation between the volume of the set of evolved states under the dephasing map (blue) and amplitude damping (orange) and the set of recovered states in terms of the dimension for $p = 0.1$. In the inset we zoom in the orange curve.}
    \label{fig:VolComp}
\end{figure}

In Fig.~\ref{fig:VolD} we depicted how the volume of the set of accessible states under the action of the channels $\Lambda_D$ (top) and $\Lambda_{AD}$ (bottom), set of \textit{evolved} states, varies under the action of dephasing and amplitude damping channels, respectively. In both cases we observe a monotonic decrease of the volume when increasing the dimension. In Fig.~\ref{fig:VolComp} we compare the volumes of the set of evolved states $V_\Lambda$ and the volume of the set of \textit{recovered} states $V_{\mathcal{P}_{\Lambda}^{\sigma}}$ obtained by applying the Petz map to the set of evolved states for a fixed noise strength $p = 0.1$. In the $y-$axis the value is limited, such that we expect values from $min = 0$ if $V_{\mathcal{P}_{\Lambda}^{\sigma}} = V_\Lambda$ to $max = 1$ if $V_{\mathcal{P}_{\Lambda}^{\sigma}} \ll V_\Lambda$. 

Fig.~\ref{fig:VolComp} brings an interesting feature to us. For the dephasing channel, the volume of the set of recovered states $V_{\mathcal{P}_{\Lambda_D}^{\sigma}}$, with fixed $\sigma(1 - 1/d)$, decreases compared to the volume of evolved states $V_{\Lambda_D}$ when increasing the dimension. Looking at Fig.~\ref{fig:distXdim}, a direct connection can be traced between the distance and the volume of accessible states. Both quantities present a monotonic increase with the dimension. For lower dimensions, the volume of accessible states for the Petz map and the dephasing channel are very similar, implying a smaller distance and consequently a better recovery. However, the same conclusion cannot be taken for the amplitude damping channel. Fixing $\sigma(0.01)$ the volumes $V_{\mathcal{P}_{\Lambda_{AD}}^{\sigma}}$ and $V_{\Lambda_{AD}}$ are quite similar for all the investigated dimensions, its relation is displayed in the inset of Fig.~\ref{fig:VolComp}. From Fig.~\ref{fig:VolComp} we cannot conclude that the performance of the Petz map will be optimal just because the aforementioned volumes are quite similar. The volumes solely assess the size of the accessible states set, overlooking the possibility of non-coincidence among them. This discrepancy becomes apparent when analyzing the amplitude damping channel, a non-unital map. The volume is not an appropriate figure of merit when dealing with non-unital maps, Eq. (\ref{eq:volume}) fails to account the translation induced by the map. Such maps introduce additional intricacies not captured by volume, due to their non-unital nature.

\subsection{Qubit case}

In here we aim to provide a geometrical visualization of the set of accessible states described previously. To this end, we restrict our analysis to qubits. We investigate how the set of qubit states vary under the action of a quantum channel and its corresponding Petz recovery map. For $d=2$, the Bloch vector is 3-dimensional. The state space of all qubit density operators is represented by a ball of unit radius in $\mathbb{R}^3$, called the Bloch ball. Each point $(r_x, r_y, r_z)$ of this ball corresponds to a unique state of the qubit.  

\begin{figure*}[t]
    \centering
\includegraphics[width=0.99\textwidth]{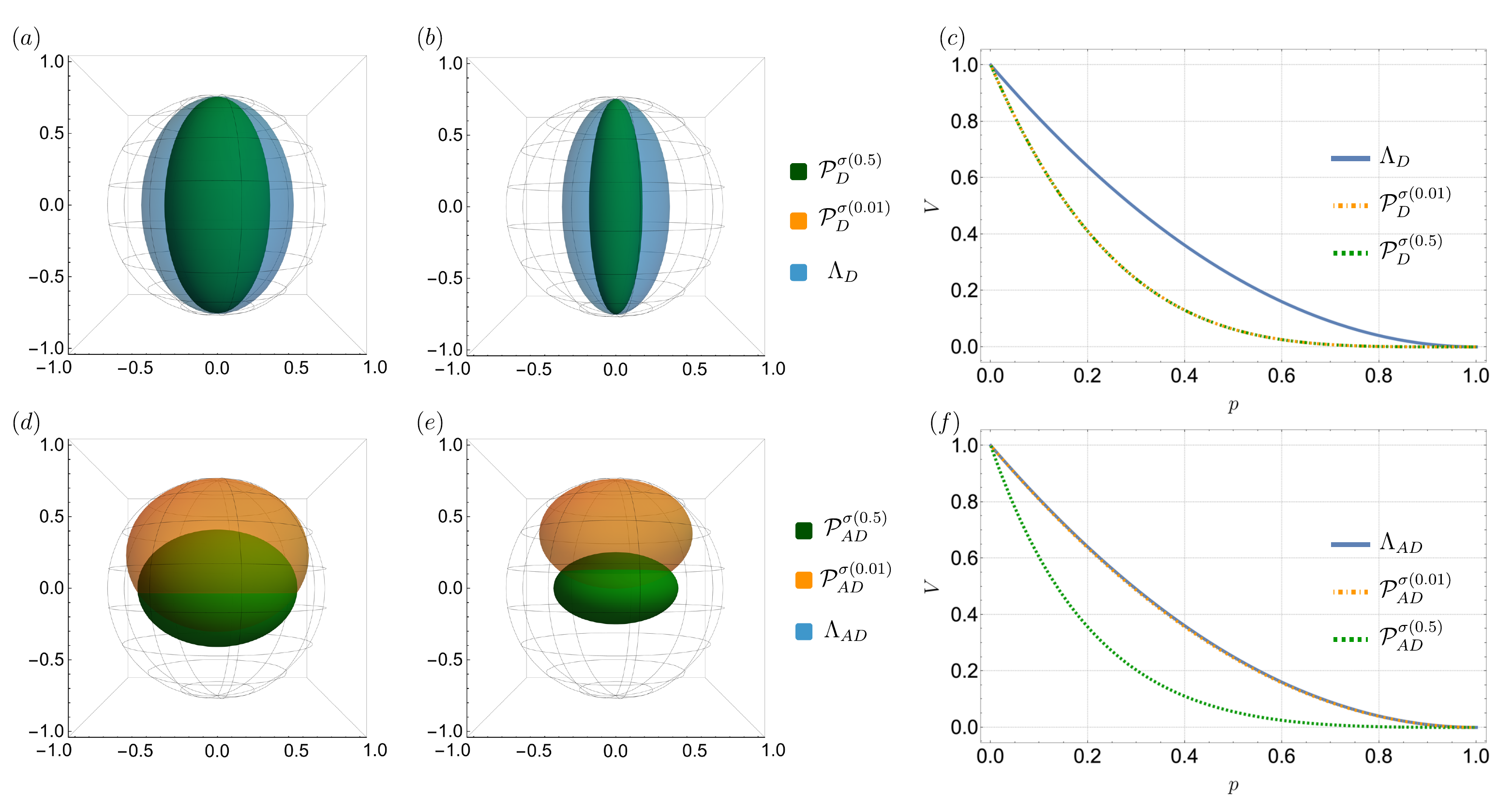}
    \caption{$(a)$ Set of accessible states under the action of the dephasing channel $\Lambda_D$ (blue ball) and the corresponding Petz recovery map for  $\sigma(0.5)$ (green ball) and $\sigma(0.01)$ (orange ball) with $p = 0.3$. The sets of accessible states for the Petz map (green and orange ball) coincide for both choices of reference states. $(b)$ Now fixing $p = 0.5$. $(c)$ Volume $V$ of accessible states for the dephasing channel $\Lambda_D$ (blue line) and its corresponding Petz recovery map $\mathcal{P}_{D}^{\sigma(0.01)}$ (orange dashed-dotted line) and $\mathcal{P}_{D}^{\sigma(0.5)}$ (green dashed line) in terms of the noise strength $p$. $(d)$ Set of accessible states under the action of the amplitude-damping channel $\Lambda_{AD}$ (blue ball) and the corresponding Petz recovery map for  $\sigma(0.5)$ (green ball) and $\sigma(0.01)$ (orange ball) with $p = 0.3$. The sets of accessible states for the $\Lambda_{AD}$ and $\mathcal{P}_{AD}^{\sigma(0.01)}$ (blue and orange ball) coincide. $(e)$ Now fixing $p = 0.5$. $(f)$ Volume $V$ of accessible states for the amplitude-damping channel $\Lambda_{AD}$ (blue line) and its corresponding Petz recovery map $\mathcal{P}_{AD}^{\sigma(0.01)}$ (orange dashed-dotted line) and $\mathcal{P}_{AD}^{\sigma(0.5)}$ (green dashed line) in terms of the noise strength $p$.}
    \label{fig:Bloch_volume}
\end{figure*}

It is known that for the qubit dephasing channel, states lying on the $z$-axis of the Bloch ball are invariant. These are the well-known fixed points of the map, $\Lambda(\rho) = \rho$. The set of accessible states under the dephasing map is then defined by the set of vectors of the form $\vec{r'} = [(1-p)r_x, (1-p)r_y, r_z]^T$. To obtain the set of accessible states for the Petz recovery map, we restrict the parametrization for the reference state given by Eq.~(\ref{eq:sigma_dim}) for $d = 2$ 
\begin{equation}
\label{eq:sigma}
\sigma(\epsilon)=(1-\epsilon) |0\rangle\langle 0| + \epsilon |1\rangle\langle 1|,
\end{equation}
with $\epsilon\in [0,1]$, a parameter that determines where the reference state lies on the $z$-axis. Choosing $\sigma$ as in Eq. (\ref{eq:sigma}), we build a Petz map $\mathcal{P}_{D}^{\sigma}$ for the dephasing channel, such that the recovered state is given by $\rho_\mathcal{P} = \mathcal{P}^{\sigma}_{D}(\Lambda_{D}(\rho))$, for any initial state $\rho$. The corresponding Bloch vector $\vec{r}_{\mathcal{P}}$ is then written as 
\begin{equation}
    \begin{split}
        \label{eq:bloch_deph}
        r_{\mathcal{P}_x} =&\, (1 + p^2 - 2 p)r_x \\
        r_{\mathcal{P}_y} =&\, (1 + p^2 - 2 p)r_y \\
        r_{\mathcal{P}_z} =&\, r_z .
    \end{split}
\end{equation}
As expected, $\vec{r}_{\mathcal{P}}$ is independent of $\epsilon$. This issue was already discussed in \cite{Lea21}; any reference state of the form of Eq. (\ref{eq:sigma}) is an optimal choice for $\mathcal{P}^{\sigma}_{D}$. 

In Fig.~\ref{fig:Bloch_volume} $(a)$ and $(b)$, we depict the set of accessible states under the dephasing map in blue and by applying the Petz recovery map to it, we obtained the set of accessible states of the Petz map in green. As expected by the symmetry of the dephasing channel, there is no distinction between the sets of recovered states of the Petz map when choosing different reference states (the green and orange balls in the plot are overlapped). However, we observe that the set of accessible states for the Petz map is always smaller than the set of accessible states for the dephasing map.

In  Fig.~\ref{fig:Bloch_volume} $(c)$, we notice how the volume $V$ of the accessible states decreases with the noise strength of the map $p$. For intermediate noise strengths the volume for $\mathcal{P}^{\sigma}_{D}$ decreases faster than for $\Lambda_{D}$. As a simple example, let us take an initial state $\rho = |+\rangle\langle +|$, with $|+\rangle = (|0\rangle + |1\rangle)/\sqrt{2}$. The evolved state under dephasing is given by $\Lambda_D(\rho)(p) = \rho'(p) = \frac{1}{2}[|0\rangle\langle 0| + (1 - p)(|0\rangle\langle 1| + |1\rangle\langle 0|) + |1\rangle\langle 1|]$. The recovered state by the Petz map is then given by $\rho_{\mathcal{P}}(p) = \frac{1}{2}[|0\rangle\langle 0| + (1 + p^2 - 2p)(|0\rangle\langle 1| + |1\rangle\langle 0|) + |1\rangle\langle 1|]$, following Eq. (\ref{eq:bloch_deph}). Choosing $p = 0.3$, as done in Fig.1, the recovered state by the Petz map is further away from the initial state than the evolved state $\rho'$. The same is observed for $p = 0.5$. Since the Petz map is CPTP, i.e. its set of accessible states cannot be bigger than the set of accessible states after applying the map, the best scenario would be to at least return $\rho_{\mathcal{P}} = \rho'$, for all the states in this set. We conclude that recoverability would imply that \textit{at least} the set of accessible states for the Petz map should coincide with the set of evolved states of the decoherence channel.

Moving to the non-unital case, under the action of the amplitude damping channel, the set of accessible states is defined by vectors of the form $\vec{r'} = [\sqrt{1-p}\,r_x, \sqrt{1-p}\,r_y, (1-p)\,rz + p \, \tau_z]^T$. Observe the shift in the $z$-direction, given by the translation component $\tau_z$. Taking the reference state of the form of Eq.~(\ref{eq:sigma}), we obtain the corresponding Petz recovered states for the amplitude damping channel $\mathcal{P}^{\sigma}_{AD}$, whose Bloch vector is 
\begin{equation}
    \begin{split}
        \label{eq:bloch_petzAD}
        r_{\mathcal{P}_x} =& \frac{\sqrt{1 - \epsilon}\sqrt{|1-p|}}{\sqrt{1 - \epsilon(1 - p)}}\,r_x ,\text{ }
        r_{\mathcal{P}_y} = \frac{\sqrt{1 - \epsilon}\sqrt{|1-p|}}{\sqrt{1 - \epsilon(1 - p)}}\,r_y \\
        r_{\mathcal{P}_z} =& \frac{1 - \epsilon +\epsilon p - p}{1 - \epsilon(1 - p)}\,r_z , \text{ }
        \tau_{\mathcal{P}_z} =  \frac{p(1 - 2\epsilon)}{1 - \epsilon(1 - p)}.
    \end{split}
\end{equation}
The Petz map is now a non-unital map, depending on the choice of $\epsilon$ and $p$, observe the dependency of $\tau_{\mathcal{P}_z}$. From Eq.~(\ref{eq:bloch_petzAD}), we see that the recovered state depends not only on the strength of the map, given by $p$, but also on the choice of the reference state, given by $\epsilon$, contrary to what was observed in Eq.~(\ref{eq:bloch_deph}) for the dephasing channel. Again we choose two specific reference states: $\sigma(0.5)$ corresponds to the maximally mixed state and $\sigma(0.01)$ is a mixed state close to the fixed point of the map $|0\rangle\langle 0|$~\cite{footnote_1}. In Fig.~\ref{fig:Bloch_volume} $(d)$ and $(e)$, we observe that the set of accessible states for $\mathcal{P}^{\sigma}_{AD}$ when $\sigma(0.5)$ (green ball)
is smaller than the one for $\sigma(0.01)$ (orange ball). The sets of accessible states for $\Lambda_{AD}$ and $\mathcal{P}^{\sigma(0.01)}_{AD}$ are superimposed, meaning that they coincide for all the points of the ball (the blue and orange balls coincide). It is also noticeable that $\mathcal{P}^{\sigma(0.01)}_{AD}$ is non-unital and its corresponding set of accessible states is shifted on the $z-$axis by $\tau_{\mathcal{P}_z} = 0.296$ and $\tau_{\mathcal{P}_z} = 0.492$ as in Figs.~\ref{fig:Bloch_volume} $(d)$ and $(e)$, respectively. However, independently of the noise strength of the map, $\mathcal{P}^{\sigma(0.5)}_{AD}$ is always a unital map, from Eq. (\ref{eq:bloch_petzAD}), we obtain $\tau_{\mathcal{P}_z} = 0$. In Fig.~\ref{fig:Bloch_volume} $(f)$, we observe how the volume of accessible states for the Petz map decreases faster by choosing $\sigma$ as the maximally mixed state. On the contrary, taking $\sigma$ as a state close to the fixed point of the map, the volume decreases at the same rate as the map itself. This results reinforce what has been discussed in the main text regarding the non-unital character of the amplitude damping map and its corresponding Petz map depending on the choice of $\sigma$.

\end{appendix}


\begin{thebibliography}{10}
\bibitem{footnote_1}
The choice $\epsilon = 0.01$ is based on the fact that the results converge for $\epsilon < 0.01$.
\end{thebibliography}
\end{document}